\documentclass[a4paper,10pt,twoside]{cpc-hepnp}

\usepackage{multicol}
\usepackage{graphicx}
\usepackage{booktabs}
\usepackage{amssymb,bm,mathrsfs,bbm,amscd}
\usepackage[tbtags]{amsmath}
\usepackage{lastpage}

\begin{document}

\fancyhead[c]{\small Chinese Physics C~~~Vol. xx, No. x (201x) xxxxxx}
\fancyfoot[C]{\small 010201-\thepage}

\footnotetext[0]{Received xx March 20xx}

\title{Nuclear Matter Incompressibility Effect on the Cross Section of Fusion Reactions with a weakly bound projectile}

\author{%
      S. A. Seyyedi$^{1}$\email{amin.seyyedi@gmail.com}
\quad H. Golnarkar $^{1}$
\quad
\quad
}
\maketitle

\address{%
$^1$ Sciences Faculty, Physics group, Payame-Noor University, P. O. Box 19395-4697, Tehran, Iran\\
 }

\begin{abstract}
Fusion reactions with a weakly bound projectile are studied using the double-folding model along with a repulsive interaction modifying term. Using this modified potential, including nuclear matter incompressibility effects, the fusion reaction cross sections and suppression parameters are calculated for $^9{Be}$ + $^{209}Bi$, $^{208}Pb$,
$^{29}Si $ and $^{27}Al$ reactions. The results show that applying these effects at energies near the Coulomb barrier improves the agreement between the calculated and experimental cross sections, and modifies the mean values of the suppression parameter.
\end{abstract}



\footnotetext[0]{\hspace*{-3mm}\raisebox{0.3ex}{$\scriptstyle\copyright$}2013
Chinese Physical Society and the Institute of High Energy Physics
of the Chinese Academy of Sciences and the Institute
of Modern Physics of the Chinese Academy of Sciences and IOP Publishing Ltd}%

\begin{multicols}{2}
{\noindent \bf {1  Introduction}}\\
\\
In recent years, there have been many theoretical and experimental studies of heavy-ion reactions with light projectiles which have low binding energy [1-5]. Such weakly bound nuclei as $^{6,7}Li $ and $^{9}Be $ are of great importance in forming super-heavy nuclei and understanding the nuclear structures. As a result of the low binding energy of such nuclei compared with tightly bound nuclei, it is interesting to follow the reaction through various channels including complete fusion (CF) and incomplete fusion (ICF). It has recently been reported that the complete fusion cross sections of such reactions in theoretical calculations where breakup is not included are about 25-30 percent greater than the experimental cross sections [6]. The absolute value of this difference is known as the suppression parameter. This parameter in the complete fusion cross section seems to originate from parallel channels of the reaction during the CF direct process, which competes with the main channel of the expected reaction. Studies in recent years have shown that the input channels of a reaction, such as total interaction potential, have a significant impact on changing this parameter. This study aims to investigate the change of rate of this parameter as a result of the application of modifying effects to the double-folding potential in  $^{9}Be $ reactions with $^{209}Bi$, $^{208}Pb$,
$^{29}Si $ and $^{27}Al$ as target nuclei, the experimental data of which have recently been presented [7-10]. These reactions are studied using the modified double-folding potential and coupled-channel method to calculate the complete cross sections. We also want to compare light target nuclei such as $^{29}Si $ and $^{27}Al$ with $^{209}Bi$, $^{208}Pb$ and explore whether or not complete fusion is also affected by the breakup process. This is an extension of our previous work [11] in which we conclude the trend of complete fusion suppression in reactions with light and heavy targets.
\\
This paper is organized as follows:  a description of the theory employed to calculate the nuclear potentials by using the double folding (DF) model and simulation of incompressibility effects is given in Sec 2. In Sec. 3, we present and discuss the effects of these corrections on the calculations of the fusion cross-sections. Section 4 gives some concluding remarks.
\\

{\noindent \bf {2   Model for $^9$Be-induced interaction potential}}\\
\\
{\noindent \bf {2.1  Basic interaction potential}}\\
\\
One of the affecting factors in heavy-ion fusion reactions is the total interaction potential. Since the Coulomb and angular parts of the potential can be calculated with a high level of accuracy, there remains only the nuclear potential to be calculated in each interaction.
There are several methods for calculating the interaction potential. Among these models, the proximity type [12-16] and Sao-Paulo [17] potentials have been mostly used for various reactions. The phenomenological base of these models motivated us to calculate the real part of the total interaction potential with the double-folding model [18]. The following integral form is used to calculate the nuclear potential:
\begin{equation} \label{1}
V(r)=\int d\textbf{r}_P \int d\textbf{r}_T
\rho_P(\textbf{r}_P)\rho_{T}(\textbf{r}_T)v_{12}(\textbf{s})
\end{equation}																						
where $\textbf{s}=\textbf{r}+\textbf{r}_P-{\textbf{r}}_T$.  In this
relation $\rho_T{(\textbf{r}_T)}$, $\rho_P{(\textbf{r}_P)}$ and
$v_{12}(s)$ are nucleon densities of the target and projectile
and the central part of the NN interaction between two nuclei respectively. \\ To reproduce the G-matrix elements
of the NN potentials in an oscillator basis the M3Y-Paris formalism is a popular choice \cite{18}. This density-independent M3Y
interaction has been used with some success in folding model
calculations of the heavy-ion potential, and its explicit form follows
below. The direct part is
\begin{equation} \label{2}
v^{M3Y}_{dir}(r)=11062\frac{\exp(-4r)}{4r}-2537\frac{\exp(-2.5r)}{2.5r}(MeV)
\end{equation}
and the zero-range pseudo-potential can be used reliably to represent the knock-on exchange part. The Paris parameterizations of
zero-range approximation for the exchange part of the interaction can be
expressed as \cite{18}:
\begin{equation} \label{3}
v^{M3Y}_{exc}(r)=-590(1-0.002\frac{E}{A}) (MeV)
\end{equation}
where $\frac{E}{A}$ is the bombarding energy per projectile nucleon.
The experimental density distributions are used both for target and
projectile. The density distribution function of the projectile
nucleus $^{9}$Be is assumed to be of the the Harmonic Oscillator
form:
\begin{equation} \label{4}
\rho_{P}(r)=\rho_0\left[1+\alpha_{P}(\frac{r}{a_P})^2\right]\exp\left(-\frac{r}{a_P}\right)^2
\end{equation}
where the coefficients $\alpha_{P}$ and $a_{P}$
are 0.611 and 1.791 respectively \cite{19}. This form of density profile offers an effective spherical shape
for the ground state of $^{9}$Be, so it is possible to neglect the
deformation effects of the nucleus and the rotational band of the ground state
of $^{9}$Be to calculate the nuclear potential. For parameterizations
of target densities, we have used the two-parameter Fermi model (2PF)
\cite{19},
\begin{equation} \label{5}
\rho_{2PF}(r)=\frac{\rho_0}{1+\exp\left(\frac{r-R_{2PF}}{a_{2PF}}\right)}
\end{equation}
The radius $R_{2PF}$ and the diffuseness $a_{2PF}$
in the above equation are listed in Table 1.

\begin{center}
\tabcaption{ \label{tab1}  Experimental density distribution parameters for target nuclei \cite{19,20}.}
\footnotesize
\begin{tabular*}{80mm}[t]{c@{\extracolsep{\fill}}cc}
\toprule Nucleus & $R_{2PF}$ (fm)   & $a_{2PF}$(fm)   \\
\hline
$^{209}Bi$ & 6.75 & 0.468  \\
$^{208}Pb$ & 6.631 & 0.505  \\
$^{29}Si $ & 3.017&  0.52      \\
$^{27}Al$ & 3.07 &0.519  \\
\bottomrule
\end{tabular*}
\vspace{0mm}
\end{center}
\vspace{0mm}
 The six-dimensional integral, Eq. 1, can be calculated using Fourier transformation in momentum space. Thereby one can reduce this integral to a product of three one-dimensional integrals using

\begin{equation}
V(k)=\int\int d\textbf{r}_P d\textbf{r}_T
\rho_P(\textbf{r}_P)\rho_{T}(\textbf{r}_T)\int v_{M3Y}(\textbf{s})e^{(-i\textbf{K}.\textbf{r})}d\textbf{s} \nonumber\\
\end{equation}
\begin{equation}
=\int\int d\textbf{r}_P d\textbf{r}_T\rho_P(\textbf{r}_P)\rho_{T}(\textbf{r}_T)\int v_{M3Y}(\textbf{s})e^{(-i\textbf{K}.(\textbf{s}+\textbf{r}_{P}-\textbf{r}_{T})}d\textbf{s} \nonumber\\
\end{equation}
\begin{equation}
=\int \rho_{P}(r_{P})e^{-i\textbf{k}.\textbf{r}_{P}}d\textbf{r}_{P}\int \rho_{T}(r_{T})e^{-i\textbf{k}.\textbf{r}_{T}}d\textbf{r}_{T}\int v_{M3Y}(\textbf{s})e^{-i\textbf{k}.\textbf{s}}d\textbf{s}\nonumber\\
\end{equation}
\begin{equation}
=\widetilde{\rho}_{P}(\textbf{k})\widetilde{\rho}_{T}(\textbf{k}) \widetilde{v}(\textbf{k})
\end{equation}

In these calculations the Fourier transformation is given by:
\begin{equation}
\widetilde{f}(k)=4\pi\int_{0}^{\infty}f(r)j_{0}(kr)r^{2}dr\\
\end{equation}
and its inverse transform is:
\begin{equation}
f(r)=1/(2\pi^{2})\int_{0}^{\infty}\widetilde{f}(k)j_{0}(kr)k^{2}dk
\end{equation}
In the calculation of total potentials the Coulomb potential is\\
\begin{eqnarray}
V_{c}=(Z_{P}Z_{T}e^{2}/{2R})(3-r^{2}/R^{2}) \;     r<R \nonumber\\
 =(Z_{P}Z_{T}e^{2}/{2R})    \;              r\geq R
\end{eqnarray}
where $R=1.2A^{1/2}_T$ and $A_{T}$ is the mass number of the target nucleus.

The process of density distributions overlapping of interaction nuclei implies that the variation of total density of colliding nuclei
can be increased up to the value of the saturation density of nuclear
matter i.e. $\rho_{0}\simeq0.17 fm^{-3}$. This non-physical effect can be modified by including the
effect of equation of state of nuclear matter in calculation of the
nuclear potential. This is explained in the following.
\\

\noindent{\bf {2.2  Modified interaction potentials}}\\

We expect that the interaction potential between target and projectile nuclei will contain an additional repulsive core in the interior region based on the Pauli Exclusion Principle, which prevents the complete overlapping of the wave functions of two fermionic systems [21]. The DF model is one of the theoretical models which uses the sudden approximation to analyze the fusion process. Therefore, based on this model, the total density in the overlapping region can be increased to values up to $\rho\simeq2\rho_{0}$. Therefore this increasing energy in the total potential at this region can be estimated by
\begin{equation}
\\ \Delta U\approx2A_P[\varepsilon(2\rho_0,\delta)-\varepsilon(\rho_0,\delta)]
\end{equation}
where $A_{P}$ is the mass number of the projectile nucleus, $\delta$ is the relative neutron excess and $\varepsilon(\rho_0,\delta)$ is the energy per nucleon as evaluated by the Thomas-Fermi equation of state for cold nuclear matter [21]. By employing a repulsive core in the N-N interaction, $v_{rep}(r)=\nu_{rep}\delta(r)$ , one can simulate this increasing energy in the DF calculation,
\\
\begin{equation}
V_{12}(r)=V_{atr}(r)+V_{rep}(r)
\end{equation}
\\
where $V_{atr}(r)$ is the attractive part of the NN interaction and
$V_{rep}(r)$ is the repulsive core which appears from the Pauli exclusion
principle. We have used the DF method, Eq.(1-8), to calculate the
attractive and repulsive terms of the nuclear part of the total potential.
In calculating the repulsive part of the nuclear potential, it is
assumed that the diffuseness parameter for the density distributions
of target and projectile are equal to $a_{rep}$. The free parameters
$a_{rep}$ and $\nu_{rep}$ are adjusted so as to meet the
following condition,
\begin{equation} \label{8}
\Delta U=V_{rep}(0)\approx\frac{A_P}{9}K
\end{equation}
where $ V_{rep}(0)$ is equal to the repulsive part of the total
potential evaluated at r = 0 by the DF integral, and K is the
incompressibility of cold nuclear matter.
 \\

\noindent{\bf {3  Calculations and results}}\\
	
In this study the total interaction potentials have been calculated in momentum space using Eq. (1-8) for $^{9}$Be+$^{209}$Bi,$^{208}$Pb,$^{29}$Si,$^{27}$Al fusion reactions. Harmonic oscillator (HO) and two-parameter Fermi (2PF) density distributions have been used in the calculations for projectile and target nucleus distribution functions. These interactions are named M3Y and depicted in Fig. 1 as dashed lines. Applying the modifications related to the nuclear matter saturation and adding the repulsion core term to the M3Y potential, the interactions are named MIP and shown by solid lines in Fig. 1 .
\begin{center}
\includegraphics[width=8cm]{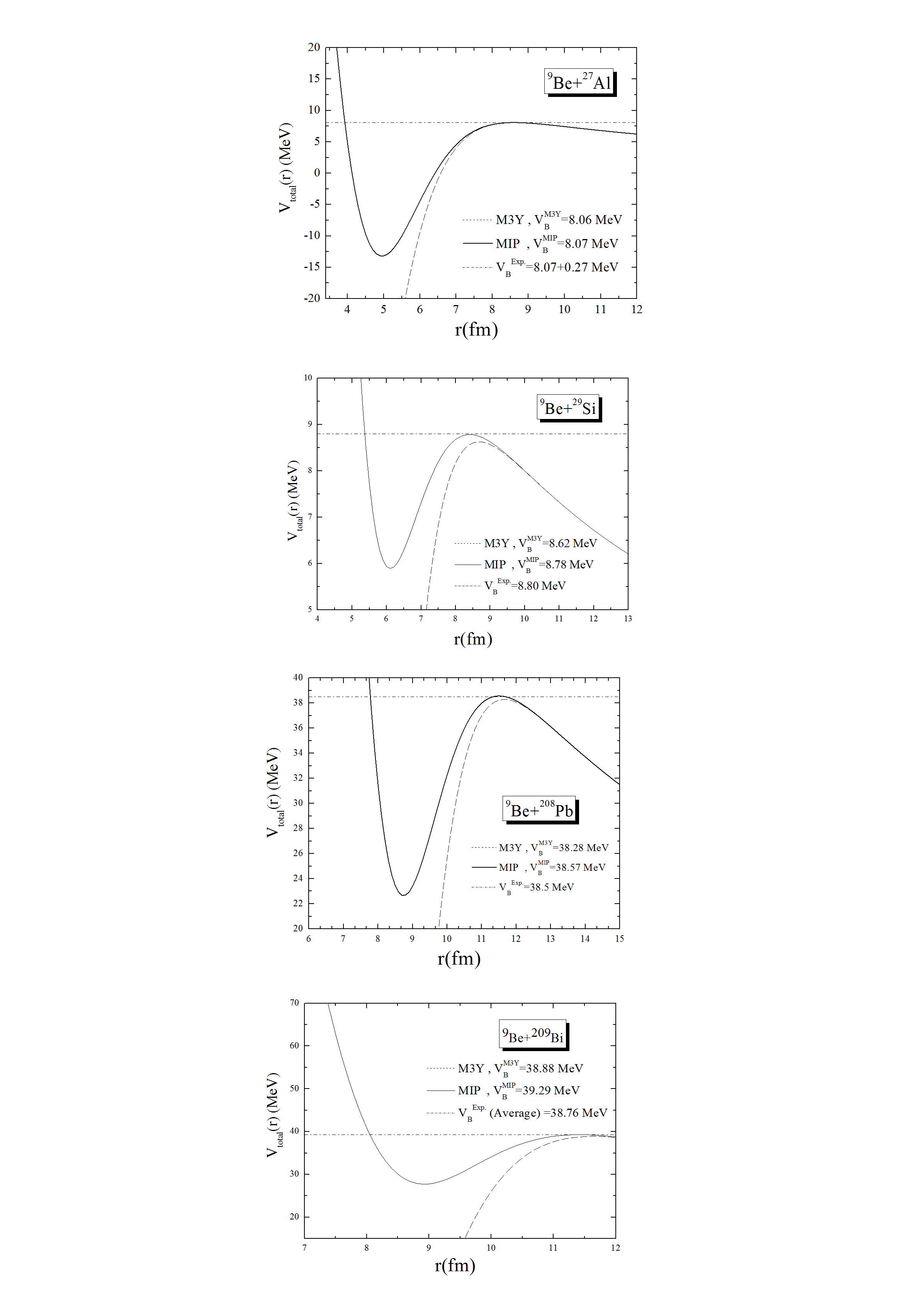}
\figcaption{\label{fig1}  Radial shape of the folded potentials for the $^{9}$Be+$^{209}$Bi,$^{208}$Pb,$^{29}$Si,$^{27}$Al fusion reactions calculated using the M3Y-Paris interaction (dashed lines) and MIP results (solid lines).}
\end{center}
As can be seen in Fig. 1 the inner regions of the M3Y potential include a very deep valley, meaning a very strong absorption in that region which is physically impossible. With the applied modifications, a valley of finite depth emerges, which seems to be necessary in the known M3Y potential.
The obtained results for the interaction potentials were used to calculate the fusion cross sections using the CCFULL coupled-channels code [22]. The calculated cross sections based on M3Y and MIP interactions are compared to experimental data for the fusion of $^{9}$Be+$^{209}$Bi,$^{208}$Pb,$^{29}$Si,$^{27}$Al in Fig. 2.
\begin{center}
\includegraphics[width=8cm]{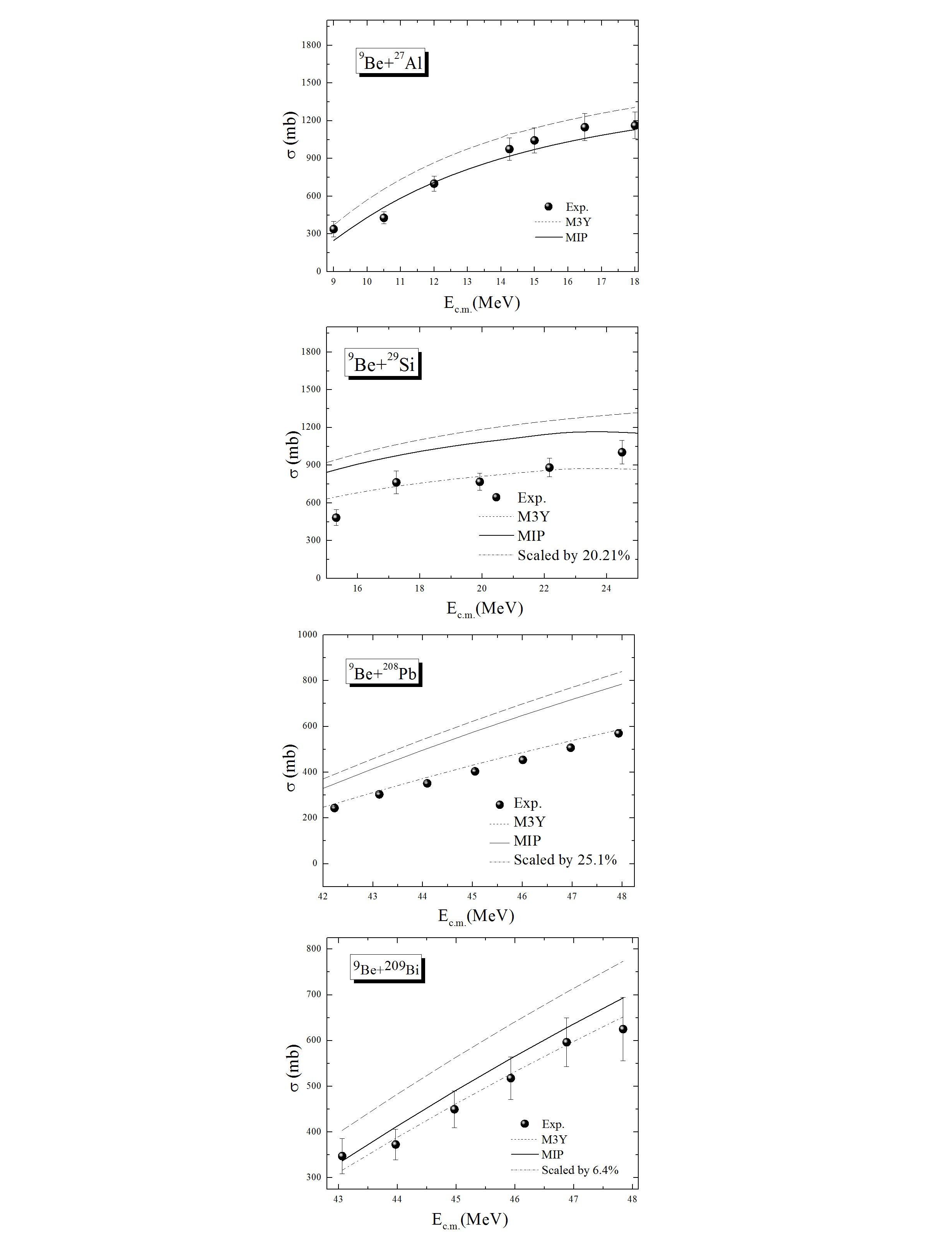}
\figcaption{\label{fig2} The fusion cross sections for the $^{9}$Be+$^{209}$Bi,$^{208}$Pb,$^{29}$Si,$^{27}$Al systems are compared to the couple-channel calculations using M3Y interaction (dashed line) and MIP interaction (solid line). The experimental data also have shown by solid circles [7-10].}
\end{center}
In Fig. 2 the dashed lines indicate the M3Y based cross sections and the solid lines indicate the MIP based cross sections. The calculated cross sections using MIP interactions, which are based on nuclear matter incompressibility, show a better agreement with experiment than calculations based on the M3Y interaction. This was achieved by adding the repulsive part of the NN interaction that is used in the calculation of DF nuclear potentials.
From these figures the existence of adding a repulsive core to the calculation of interaction potential seems to be helpful in describing weakly bound fusion systems.\\

\begin{center}
\includegraphics[width=13cm]{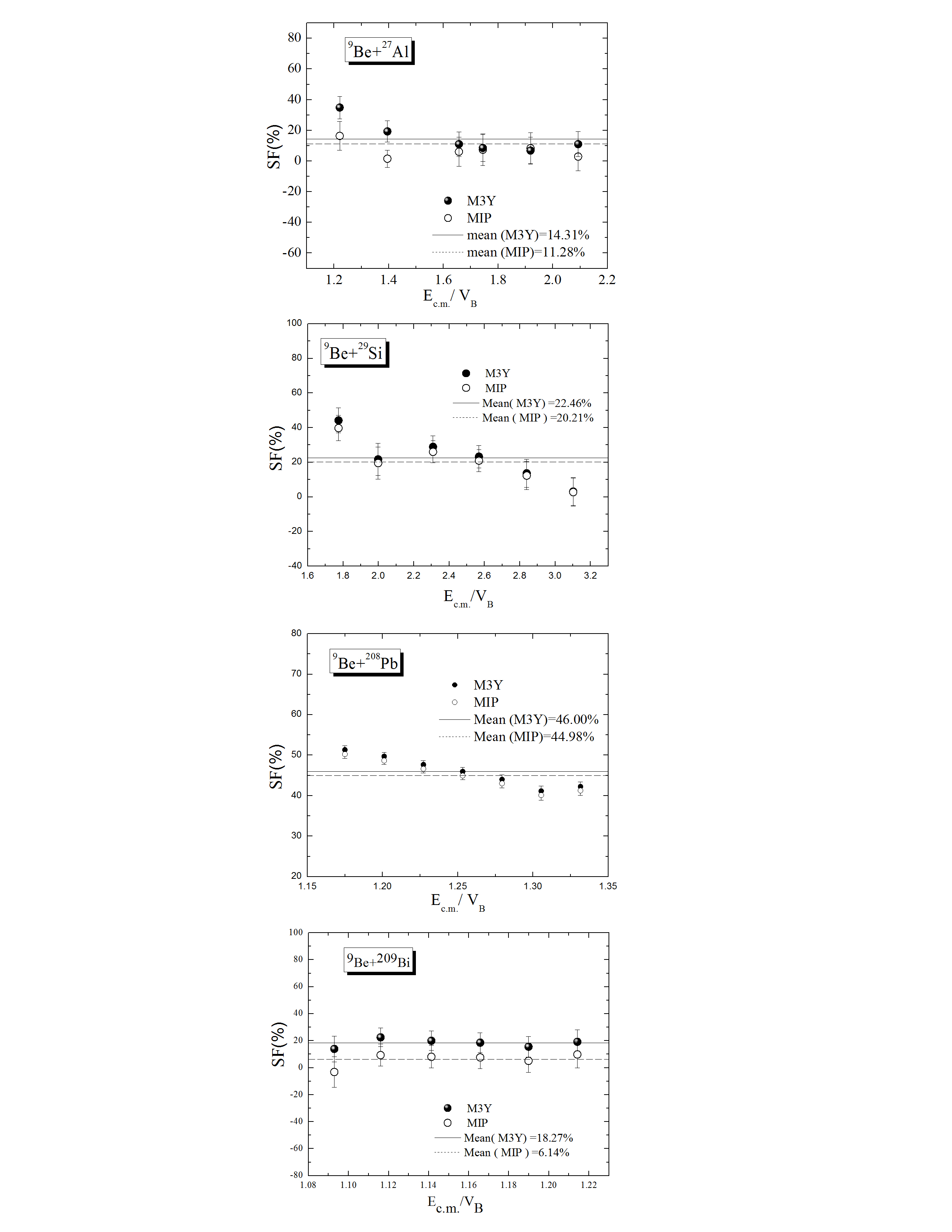}
\figcaption{\label{fig3} Suppression parameter for the $^{9}$Be+$^{209}$Bi,$^{208}$Pb,$^{29}$Si,$^{27}$Al fusion systems. The solid circles indicate the M3Y based calculations and the open circles indicate the MIP based calculations. The mean values for each reaction are indicated in the figure. }
\end{center}

A wide variety of studies on weakly bound nuclei reactions in recent years have introduced a modifying term [23,24]. These investigations
suggest the necessity of introducing the suppression parameter (SP) in the theoretical calculations. This parameter is dependent on the potential
 model used in the interaction and is defined as follows
\begin{equation}
SP=(1-\sigma_{Exp.}/\sigma_{Theor.})\times100
\end{equation}

where $\sigma_{Exp.}$ and $\sigma_{Theor.}$ are fusion cross sections obtained from the experimental measurements and theoretical calculations, respectively. Therefore, the repulsion term effect resulting from overlapping nuclear densities of the two interactive nuclei on the suppression parameter mean value is calculated using Eq. (13) for the MIP and M3Y models and displayed in Fig. 3. in terms of $E_{c.m.}/V_{B}$.

 It can be observed from this figure that the nuclear matter saturation effects can modify  the suppression
 parameter values. The changes are listed in Table 2.
\\
\tabcaption{ \label{tab2} Suppression parameter values for fusion reactions before and after the compressibility of nuclear matter is included.}
\footnotesize
\begin{tabular*}{80mm}[t]{c@{\extracolsep{\fill}}cc}
\toprule Nucleus & $SP(M3Y)$  & $SP(MIP)$  \\
\hline
$^{209}Bi$ & 18.27 & 6.14  \\
$^{208}Pb$ & 46.00 & 44.98  \\
$^{29}Si $ & 22.46&  20.20      \\
$^{27}Al$ & 14.31 &11.28  \\
\bottomrule
\end{tabular*}
\\
\\
\\
\normalsize
The differences which we see in Table 2 for the SP(M3Y) in the
$^{9}$Be+$^{209}$Bi,$^{208}$Pb, systems is ~28 percent.  Fusion reactions with $^{9}$Be,$^{6,7}$Li beams incident on $^{209}$Bi and $^{208}$Pb targets have investigated in [25-31]. Above-barrier CF cross sections for the reaction
of $^{6}$Li with $^{209}$Bi and $^{208}$Pb are in agreement with each other and both show about 34 percent for the SP value
[28,29,32]. However, the situation is not so clear for the $^{9}$Be-induced reactions, where
the difference between measured complete fusion cross sections for the $^{9}$Be+$^{209}$Bi system and the neighboring
$^{9}$Be+$^{208}$Pb system is about 25 percent [33-36]. Furthermore, discrepancies exist even between the different sets of complete fusion measurements for $^{9}$Be+$^{209}$Bi, which were carried out in different laboratories [27,37]. The large difference in above-barrier CF cross sections between reactions with $^{209}$Bi and $^{208}$Pb targets is discussed in [38], which attributes the current problem to normalization for one or more measurements.
\\
\\
\\
{4 CONCLUSIONS}\\

In summary, we have modified explicitly some generalized and realistic calculation processes for the double folding model including nuclear matter incompressibility. The modified program is applied to calculate the real part of the interaction potential for several weakly bound fusion systems. This study also examines the effects of modeling the repulsion term in the nuclear potential on the fusion cross sections and the effect of the suppression parameter of nuclear fusion reactions with a weakly bound projectile. The studies have been carried out in this case on $^9{Be}$ + $^{209}Bi$, $^{208}Pb$,
$^{29}Si $ and $^{27}Al$  reactions. The results show that applying the nuclear matter incompressibility effects to the nuclear potential calculation limits the nuclear potential depth in the inner regions of the ion-ion distance. As shown in Fig. 2, this improves the agreement between the calculated and experimental cross sections. The applied modifications also modify the mean value of the suppression parameter, which is displayed in Fig. 3 for $^9{Be}$ + $^{209}Bi$, $^{208}Pb$,
$^{29}Si $ and $^{27}Al$  reactions.
\\

\end{multicols}

\clearpage


\begin{thebibliography}{99}

\bibitem{1} Canto L F, Gomes P R S, Donangelo R et al, Phys. Rep., 2006, 424:1
\bibitem{2} Hinde D J, Dasgupta M, Phys. Rev. C, 2010, 81:064611
\bibitem{3} Pradhan M K, Mukherjee A, Basu P et al, Phys. Rev. C, 2011, 83:064606
\bibitem{4} Esbensen H, Phys. Rev. C, 2010, 81:034606
\bibitem{5} Otomar D R, Lubian J, Gomes P R S et al, Phys. Rev. C, 2009, 80:034614
\bibitem{6} Gomes P R S, Linares R, Lubian J et al, Phys. Rev. C, 2011, 84:014615
\bibitem{7} Dasgupta M, Hinde D J, Sheehy S L et al, Phys. Rev. C, 2010, 81:24608
\bibitem{8} Figueira M C S, Szanto E M, Anjos R M et al, Nuclear Physics A, 1993, 561:453
\bibitem{9} Marti G V, Gomes P R S, Rodriguez M D et al, Phys. Rev. C, 2005, 71:027602
\bibitem{10} Liu Z H, Signorini C, Mazzocco M et al, European Physical Journal A, 2005, 26:73
\bibitem{11} Ghodsi O N, Seyyedi S A, Mohammadi S, Mod. Phys. Lett. A, 2012, 27:1250133
\bibitem{12} Dutt I, Puri R K, Phys. Rev. C, 2010, 81:044615
\bibitem{13} Dutt I, Puri R K, Phys. Rev. C, 2010, 81:064608
\bibitem{14} Dutt I, Puri R K, Phys. Rev. C, 2010, 81:064609
\bibitem{15} Puri R K, Chattopadhyay P, Gupta R K, Phys. Rev. C, 1991, 43:315
\bibitem{16} Puri R K, Gupta R K, Phys. Rev. C, 1992, 45:1837
\bibitem{17} Chamon L C, Nucl. Phys. A, 2007, 787:198
\bibitem{18} Satchler G R, Love W G, Phys. Rep., 1979, 55:183
\bibitem{19} De Vries H,  De Jager C W, De Vries C, At. Data. Nucl. Data Tables, 1987, 36:495536
\bibitem{20} Gontchar I, Hinde D J, Dasgupta M et al, Phys. Rev. C, 2004, 69:024610
\bibitem{21} Misicu S, Esbensen H, Phys. Rev. C, 2007, 75:034606
\bibitem{22} Hagino K, Rowley N, Kruppa A T, Comp. Phys. Comm., 1999, 123:143
\bibitem{23} Wang B, Zhao W, Gomes P R S et al, Phys. Rev. C, 2014, 90:034612
\bibitem{24} Gomes P R S, Canto L F, Lubian J et al, J. Phys. G:Nucl. Part. Phys., 2012,  39:11
\bibitem{25} Freiesleben H, Britt H C, Birkelund J et al, Phys. Rev. C 1974, 10:245
\bibitem{26} Dasgupta M, Hinde D J, Butt R D et al, Phys. Rev. Lett., 1999, 82:1395
\bibitem{27} Signorini C, Liu Z H, Li Z C et al, Eur. Phys. J. A, 1999, 5:7
\bibitem{28} Dasgupta M, Hinde D J, Hagino K, Phys. Rev. C, 2002, 66:041602 
\bibitem{29} Wu Y W, Liu Z H,  Lin C J et al, Phys. Rev. C, 2003, 68:044605
\bibitem{30} Dasgupta M, Gomes P R S, Hinde D J et al, Phys. Rev. C, 2004, 70:024606 
\bibitem{31} Signorini C, Liu Z H, Yoshida A, Eur. Phys. J. A, 1998, 2:227 
\bibitem{32} Rusek K, Alamanos N, Keeley N et al, Phys. Rev. C, 2004, 70:014603
\bibitem{33} Dasgupta M, Hinde D J, Butt R D et al, Phys. Rev. Lett., 1999, 82:1395
\bibitem{34} Signorini C, Liu Z H, Li Z C et al, Eur. Phys. J. A, 1999, 5:7 
\bibitem{35} Signorini C, Glodariu T, Liu Z H,  Prog. Theor. Phys. Suppl., 2004, 154:272 
\bibitem{36} Liu Z H, Signorini C, Mazzocco M at al, Eur. Phys. J. A, 2005, 26:73 
\bibitem{37} Yoshida A, Signorini C, Fukuda T et al, Phys. Lett. B, 1996, 389:457 
\bibitem{38} Dasgupta M, Hinde D J, Sheehy S L et al, Phys. Rev. C, 2010, 81:024608 

\end{thebibliography}
\end{document}